# On the Photon Motion Near a Five-Dimensional Schwarzschild Black Hole

Husin Alatas[1,3,#], Siti A. Nuraeni[1], Ilma L. Saptiani[1], Bobby E. Gunara[2,3]

[1])*Theoretical Physics Division, Department of Physics, IPB University, Jl. Meranti, Kampus IPB Darmaga, Bogor 16680, Indonesia*

[2])*Theoretical High Energy Physics Research Group, Faculty of Mathematics and Natural Sciences, Institut Teknologi Bandung, Jl. Ganesha 10, Bandung 40132, Indonesia*

[3])*Indonesia Center for Theoretical and Mathematical Physics (ICTMP)*

[#])*Corresponding author email: alatas@apps.ipb.ac.id*

## Abstract

In this report, we discuss the dynamics of photons in a five-dimensional Schwarzschild black hole. We find that the corresponding Eddington-Finkelstein-like coordinate admits two time-coordinate singularities for a particular condition. This condition is due to the existence of a non-zero angular momentum related to the extra spatial dimension. By introducing a new Regge-Wheeler tortoise coordinate, both singularities can be removed from the Eddington-Finkelstein-like coordinate. It is identified that the additional singularity is related to the photon sphere in a hyperplane, and we analyze it using a dynamical system approach. Simulation results of photon motion based on numerical calculation are presented. It is suggested that the related bending motion characteristic can probably be considered as a signature for probing the existence of the extra spatial dimension.

# 1. Introduction

The success of reconstructing the M87* black hole image by the Event Horizon Telescope (EHT) [1] has been undoubtedly one of the triumphs of general relativity theory. Historically, back to 1916, K. Schwarzschild has proposed a spherically symmetric metric solution of Einstein field equation that contains singularity at the origin [2, 3]. While in the same year, J. Droste has also found independently the same metric in which the chosen coordinate system has indicated the existence of a static black hole nowadays called the Schwarzschild black hole [4]. Later, other black-hole solutions have been found in numerous metric forms, such as the Kerr metric of spinning black hole, Kerr-Newman metric of a charge-spinning black hole, and many others [5-9]. Physically, a black hole is an object with an enormous mass density that occupied a tiny space region where even light cannot escape from its strong gravitational pull when approaching the so-called event horizon [3], a boundary that distinguishes the interior and exterior regions of a black hole. According to the cosmic censorship conjecture [10, 11], the singularity hides inside the horizon of a black hole. Nowadays, it is well understood that the gravitational collapse during the event of the supernovae of a massive star is responsible for the occurrence of a black hole [12].

Over the past century, the Schwarzschild black hole has been intensively studied in the mathematical context, e.g. [13-17], although most likely, a static Schwarzschild black hole might never exist. Nevertheless, as a simple mathematical model, it is important to understand the nature of a black hole and its strong gravitational field behavior. In the meantime, it is well known that gravitational interaction is the weakest force among the other forces in the universe, such that it has been a great puzzle to physics until now on how this interaction shapes our present universe [16]. Many theoretical physicists have conducted efforts to understand the physical explanation of this phenomenon. One of the proposals to explain the phenomenon is the introduction of extra spatial dimensions, in which the gravitational interaction is suspected can leak into it while others do not [19-21].

The first attempt to extend the spatial dimension was conducted by T. Kaluza and O. Klein [22] in order to unify the gravitation and electromagnetic interactions into a single theory. Although the theory failed to reach its goal, the idea of extra spatial dimensions inspires many physicists. For instance, in the Dvali-Gabadadze-Porrati (DGP) braneworld theory [19, 20], it is assumed that our existing universe is actually embedded in a bulk universe with extra spatial dimensions.

In the meantime, the existence of higher-dimensional black holes has been studied and reported by many theorists, including that of related to the Schwarzschild black hole [23-27]. One of the most remarkable findings is the predicted existence of a black ring, in which the related singularity is in the form of $S^1 \times S^{D-3}$ topology, where $D = N+1$ is the number of dimensions [28], with $N$ is the number of spatial dimensions. In the meantime, one of the exciting features of a black hole is related to the dynamical motion of particles around it. For a $(3+1)-$dimensional black hole, the dynamics of massless and massive particles have been discussed intensively such as the motion of a massless photon around the static [29, 30] and spinning black holes [30, 31], and also the motion of massive particles around a charge-static/rotating black hole due the presence of quintessence and magnetic field [32, 33]. On the other hand, the dynamics of the massless and massive particles near a black hole with higher dimension has also been discussed [27, 34-35].

In this article, we report the results of our recent study on the photon motion around a Schwarzschild black hole in $(4+1)-$dimension. Inspired, for instance, by the DGP braneworld theory [19, 20], one can assume that our existing $(3+1)-$dimension universe is actually embedded in a $(4+1)-$dimension bulk universe whereas the related extra spatial dimension is hidden to us. Next, by considering an additional spatial dimension, while preserving the spherical symmetry, we found that the representation of our proposed metric in terms of Eddington-Finkelstein-like coordinate [36] leads to the occurrence of additional time-coordinate singularity in the associated ingoing and outgoing photon motion in a specific condition of the extra spatial dimension related angular momentum. Using a dynamical system approach, we show that the additional time-coordinate singularity is nothing but related to the photon sphere in a hyperplane. We also studied the escaping photons bending characteristics near the related black hole numerically, which can probably be used to probe the corresponding extra spatial dimension existence. To find a black hole, detecting the presence of escaping photons is an important aspect [1].

We organized this report as follows. In section 2 we discuss the corresponding metric of $(4+1)$-dimension Schwarzschild black hole, followed by a discussion on the photon trajectories and the associated Eddington-Finkelstein-like coordinate in section 3. We discuss the photon trajectories with zero and non-zero extra spatial dimension angular momentum in section 4. Conclusion of this report is given in section 5. From our perspective, the discussion conducted in this report is helpful to motivate the early-year graduate students to explore the physics formulation in a higher-dimensional space-time based on general relativity.

## 2. $(4+1)$ – Dimensional Schwarzschild Black Hole

Let us consider a static $(4+1)$-spacetime geometry that admits a rotational isometry on a 4-dimensional compact manifold. This kind of spacetime is nothing but the $(4+1)$-dimensional Schwarzschild spacetime whose coordinates are given by $(t, r, \theta, \phi, \alpha)$, denoting respectively the time-coordinate, radial-coordinate, and the three angular-coordinates. In this coordinate system, the Schwarzschild black hole solution of the vacuum Einstein equation is given by the Tangherlini-Schwarzschild-Droste line element [23], which can be written in the following form in terms of natural unit with $G = c = 1$:

$$ds^2 = g_{\mu\nu}dx^\mu dx^\nu = \left(1 - \frac{2M}{r^2}\right)dt^2 - \left(1 - \frac{2M}{r^2}\right)^{-1} dr^2 - r^2 d\Omega_3^2 \qquad (1)$$

where $M$ is the mass of a massive object, $d\Omega_3^2 = \left[\cos^2\alpha\left(d\theta^2 + \sin^2\theta d\phi^2\right) + d\alpha^2\right]$, and the metric can be written as follows:

$$g_{\mu\nu} = \text{diag}\left[-\left(1-\frac{2M}{r^2}\right),\left(1-\frac{2M}{r^2}\right)^{-1},-r^2\cos^2\alpha,-r^2\cos^2\alpha\sin^2\theta,-r^2\right] \quad (2)$$

As mentioned earlier, here we assumed that the $\alpha$-coordinate lies in the $(4+1)$-dimension bulk universe, outside our existing $(3+1)$-dimension universe. It is readily seen that the metric (2) admits an event horizon with radius $r_{EH} = \sqrt{2M}$.

Let us first discuss the geodesic equation: $\ddot{x}^\mu + \Gamma^\mu_{\nu\rho}\dot{x}^\nu\dot{x}^\rho = 0$, with $\dot{x}^\mu \equiv dx^\mu/d\sigma$, where $\sigma$ denotes an affine-parameter along the null geodesic curve of a zero-mass particle, i.e., photon, by considering the associated Lagrangian: $L = g_{\mu\nu}\dot{x}^\mu\dot{x}^\nu$ which is explicitly given by:

$$L = \left(1-\frac{2M}{r^2}\right)\dot{t}^2 - \left(1-\frac{2M}{r^2}\right)^{-1}\dot{r}^2 - r^2\left[\cos^2\alpha\left(\dot{\theta}^2 + \sin^2\theta\dot{\phi}^2\right) + \dot{\alpha}^2\right] \quad (2)$$

Applying the Lagrange equation: $d\left(\partial L/\partial \dot{x}^\mu\right)/d\sigma = \partial L/\partial x^\mu$ and after a straightforward calculation one finds the following set of differential equations:

$$\left(1-\frac{2M}{r^2}\right)\dot{t} = k \quad (3)$$

$$\left(1-\frac{2M}{r^2}\right)\dot{t}^2 - \left(1-\frac{2M}{r^2}\right)^{-1}\dot{r}^2 = r^2\left[\cos^2\alpha\dot{\phi}^2 + \dot{\alpha}^2\right] \quad (4)$$

$$r^2\dot{\phi} = \frac{h}{\cos^2\alpha} \quad (5)$$

$$\ddot{\alpha} + \frac{2}{r}\dot{r}\dot{\alpha} = \sin\alpha\cos\alpha\,\dot{\phi}^2 \quad (6)$$

Here, $k$ and $h$ are free parameters. The parameter $h$ describes the $\phi$-angular momentum related parameter of a moving particle. In later discussion it will be shown that the $k$ parameter is related to the "energy" of the moving particle. In our calculation, we have considered $L = g_{\mu\nu}\dot{x}^\mu \dot{x}^\nu = 0$ for a zero-mass particle, and without loss of generality, we only consider the geodesic of an object moving at hypersurface with $\theta = \pi/2$.

## 3. Photon Radial Motion and Eddington-Finkelstein-like Coordinate

Inspired by the DGP braneworld scenario [19, 20], which we have implicitly assumed, here we define the photon radial motion as a motion with zero $\phi-$ angular momentum since the extra spatial dimension, which is represented by $\alpha-$ coordinate can be considered to lie in the $(4+1)-$ dimensional bulk universe, which is outside the $(3+1)-$ dimensional presence observed universe.

Having Eq. (4) and (6) at hand, we are now able to calculate the null geodesic of a photon with zero-rest mass. We consider first the case with $\dot{\phi} = 0$, such that:

$$\left(1 - \frac{2M}{r^2}\right)\dot{t}^2 - \left(1 - \frac{2M}{r^2}\right)^{-1}\dot{r}^2 = r^2 \dot{\alpha}^2 \tag{7}$$

$$\ddot{\alpha} + \frac{2}{r}\dot{r}\dot{\alpha} = 0 \tag{8}$$

Remarkably, Eq. (8) can be written in the following form:

$$\frac{1}{r^2}\frac{d}{d\tau}\left(r^2 \dot{\alpha}\right) = 0 \tag{9}$$

such that

$$r^2 \dot{\alpha} = \gamma \tag{10}$$

where the integration constant $\gamma$ is a newly introduced parameter describing the so called $\alpha-$ angular momentum parameter. Next, by considering Eq. (3) one can write:

$$\dot{\alpha} = \frac{\beta}{r^2}\left(1 - \frac{2M}{r^2}\right)\dot{t} \tag{11}$$

with $\beta = \gamma/k$. Substituting Eq. (11) into Eq. (7) one finds:

$$\frac{dt}{dr} = \pm \frac{1}{\left(1 - \frac{2M}{r^2}\right)\sqrt{1 - \frac{\beta}{r^2}\left(1 - \frac{2M}{r^2}\right)}} \qquad (12)$$

with $+(-)$ sign is related to the outgoing (ingoing) photon radial trajectory. Obviously, the choice of $\beta = 0$ or $\beta \neq 0$ will leads to different behavior, where for the first case, with zero $\alpha$ – angular momentum, is related to a pure radial motion of the photon, while the second case is related to a radial motion with non-zero $\alpha$ – angular momentum.

Let us first consider the case of $\beta = 0$. Integrating the Eq. (12) for this case leads to:

$$t = -r + \frac{\sqrt{2M}}{2} \ln\left|\frac{r + \sqrt{2M}}{r - \sqrt{2M}}\right| + v \qquad (13)$$

for ingoing photon and

$$t = r - \frac{\sqrt{2M}}{2} \ln\left|\frac{r + \sqrt{2M}}{r - \sqrt{2M}}\right| + w \qquad (14)$$

for outgoing photon, where $v$ and $w$ are integration constants. Clearly, for this case the related radial trajectory admits of what we call as the "time-coordinate singularity" at $r = r_{EH} = \sqrt{2M}$, in which $\lim_{r \to \sqrt{2M}} dt/dr \to \infty$ [see Eq. (12)]. This result is similar to what was reported in ref. [25].

Next, for the general case with $\beta \neq 0$, the condition $1 - \beta(1 - 2M/r^2)/r^2 = 0$ in Eq. (12) leads us to the following two critical radii:

$$r_\pm^2 = \frac{\beta \pm \sqrt{\beta(\beta - 8M)}}{2} \qquad (15)$$

In principle, it is readily seen that the equations for the ingoing and outgoing photon with non-zero $\alpha$ – angular momentum, can be found by integrating Eq. (12). But it seems it is reasonable to think that the related integral is very difficult to solve, which involves complicated elliptic function.

However, for a systematic analysis, before we discuss the general case of $\beta$, it is readily seen that by choosing a condition $\beta = 8M$, the expression on the r.h.s of Eq. (12) can be recast into the following simple form:

$$\frac{dt}{dr} = \pm \frac{1}{\left(1-\frac{2M}{r^2}\right)\left(1-\frac{4M}{r^2}\right)} \quad (16)$$

exhibiting two time-coordinate singularities where $\lim_{r \to r_S} dt/dr \to \infty$ with $r_S = r_{EH}$ and $2\sqrt{M}$. Before we proceed further, it is useful to construct from Eq. (11) and (16) the following differential equation:

$$\frac{d\alpha_{\pm}}{dr} = \pm \frac{8M}{r^2\left(1 - 4M/r^2\right)} \quad (17)$$

with the following solution:

$$\alpha_{\pm}(r) = 4\sqrt{M} \ln\left|\frac{r \mp 2\sqrt{M}}{r \pm 2\sqrt{M}}\right| \quad (18)$$

Note that we will discuss further the general case of $\beta$ in section 4. Next, by integrating Eq. (12) one finds:

$$t = -r - \frac{\sqrt{2M}}{2} \ln\left|\frac{r+\sqrt{2M}}{r-\sqrt{2M}}\right| + \frac{\alpha_{-}(r)}{2} + p \quad (19)$$

for an ingoing photon and

$$t = r + \frac{\sqrt{2M}}{2} \ln\left|\frac{r+\sqrt{2M}}{r-\sqrt{2M}}\right| + \frac{\alpha_{+}(r)}{2} + q \quad (20)$$

for an outgoing photon. Similar to the previous case, the parameters $p$ and $q$ are the related integration constants. Compared to the previous case, it is readily seen from Eq. (19) and (20) that in addition to the first time-coordinate singularity at $r = r_{EH}$ second time-coordinate singularity also occurs at $r = 2\sqrt{M}$ which is not found in the metric (2). Note that this second time-coordinate singularity is related to the ill-defined condition of $\alpha(r)$ at $r = 2\sqrt{M}$ as shown by Eq. (18).

Depicted in Fig. 1 is the examples of the related ingoing and outgoing photon motions, showing that there are two time-coordinate singularities where the physical interpretation of the second time-coordinate singularity will be discussed in the section 4. In the following, we discuss the formulation of Eddington-Finkelstein and Eddington-Finkelstein-like coordinates based on the photon motions with $\beta = 0$ and $\beta = 8M$, respectively.

Consider first the case with $\beta = 0$. To construct the corresponding Eddington-Finkelstein coordinate for the ingoing photon we define the integration constant $v$ in Eq. (13) as a new coordinate i.e.:

$$v = t + r - \frac{\sqrt{2M}}{2} \ln\left|\frac{r + \sqrt{2M}}{r - \sqrt{2M}}\right| \qquad (21)$$

Differentiating the above coordinate, one finds:

$$dv = dt + \frac{dr}{\left(1 - \frac{2M}{r^2}\right)} \qquad (22)$$

Inserting $dt$ into metric (2) yields the following Eddington-Finkelstein coordinate in terms of ingoing photon radial trajectory:

$$ds^2 = \left(1 - \frac{2M}{r^2}\right)dv^2 - 2dvdr - r^2 d\Omega_3^2 \qquad (23)$$

Based on similar procedure, the Eddington-Finkelstein coordinate in terms of outgoing photon radial trajectory is given as follows:

$$ds^2 = \left(1 - \frac{2M}{r^2}\right)dw^2 + 2dwdr - r^2 d\Omega_3^2 \qquad (24)$$

As expected, all these results are similar to the Eddington-Finkelstein of $(3+1)$-dimensional Schwarzschild black hole.

For the ingoing photon case with $\beta = 8M$, we can construct an Eddington-Finkelstein-like coordinate for the associated motion with non-zero $\alpha$ – angular momentum, namely by defining $p$ in Eq. (19) as the related new coordinate i.e.:

$$p = t + r + \frac{\sqrt{2M}}{2}\ln\left|\frac{r+\sqrt{2M}}{r-\sqrt{2M}}\right| - \frac{\alpha_-(r)}{2} \tag{25}$$

Differentiating the above coordinate, one finds:

$$dp = dt + \frac{dr}{\left(1-\frac{2M}{r^2}\right)\left(1-\frac{4M}{r^2}\right)} \tag{26}$$

Inserting $dt$ into (1) yields the following Eddington-Finkelstein-like coordinate:

$$ds^2 = \left(1-\frac{2M}{r^2}\right)dp^2 - 2\left(1-\frac{4M}{r^2}\right)^{-1}dpdr - \left(1-\frac{2M}{r^2}\right)^{-1}\left[1-\left(1-\frac{4M}{r^2}\right)^{-2}\right]dr^2 - r^2 d\Omega_3^2 \tag{27}$$

which contains the two ill-defined points related to the time-coordinate singularities, namely at $r = \sqrt{2M}$ and $2\sqrt{M}$. We can remove these coordinate singularities by introducing a new Regge-Wheeler (RW) tortoise radial coordinate $r_{*,-}$:

$$r_{*,-} = \int \frac{dr}{1-\frac{4M}{r^2}} = r - \frac{\alpha_-(r)}{4} \tag{28}$$

such that

$$ds^2 = \left(1-\frac{2M}{r^2}\right)dp^2 - 2dpdr_{*,-} - \frac{8M}{r^2}\left(1-\frac{4M}{r^2}\right)^2 dr_{*,-}^2 - r^2 d\Omega_3^2 \tag{29}$$

Meanwhile, for the outgoing photon case with the new coordinate [see Eq. (20)]

$$q = t - r + \frac{\sqrt{2M}}{2}\ln\left|\frac{r+\sqrt{2M}}{r-\sqrt{2M}}\right| - \frac{\alpha_+(r)}{2} \tag{30}$$

one finds the RW-tortoise incorporated Eddington-Finkelstein-like coordinate as follows:

$$ds^2 = \left(1 - \frac{2M}{r^2}\right) dq^2 + 2 dq\, dr_{*,+} - \frac{8M}{r^2}\left(1 - \frac{4M}{r^2}\right)^2 dr_{*,+}^2 - r^2 d\Omega_3^2 \tag{31}$$

where the associated RW-tortoise coordinate $r_{*,+}$ is given by:

$$r_{*,+} = \int \frac{dr}{\frac{4M}{r^2} - 1} = -\left[r + \frac{\alpha_+(r)}{4}\right] \tag{32}$$

It is readily seen that all these RW-tortoise incorporated Eddington-Finkelstein-like coordinates are free from coordinate singularities. It is important to keep in mind that, in principle, the radial coordinate $r$ defined in metrics (29) and (31), is a function of the related tortoise coordinates, i.e., $r \equiv r(r_{*,-})$ and $r \equiv r(r_{*,+})$, respectively.

## 4. Photon Motions with Non-Zero Angular Momentums

### 4.1. Photon Sphere

The unclear point of our finding is the occurrence of the second time-coordinate singularity at $r = 2\sqrt{M}$, for the case $\beta = 8M$, that does not exist in the associated $(4+1)$-dimensional Schwarzschild metric (2). This time-coordinate singularity emerged from the dynamics of the additional spatial coordinate, represented by $\alpha$ function as shown by Eq. (11).

To interpret the physical meaning of the corresponding second time-coordinate singularity, we investigate the photon circular motion namely by considering the related Newtonian-like orbit equation, which can be used to determine the radius of photon circular motion in a so-called "photon sphere" $r_{PS}$ [27]. To derive the associated Newtonian-like orbit equation, we first consider the case of $\beta = 0$ with zero $\alpha$-angular momentum: $r^2 \dot{\alpha} = 0$. From Eq. (3)-(5), we obtain the following "energy' differential equation:

$$\frac{1}{2}\dot{r}^2 + V_{eff} = E \tag{33}$$

while from Eq. (6) it is readily seen that we have to set the corresponding motion is confined in a hyperplane with $\theta = \pi/2$ and $\alpha = 0$. Here, the constant parameter $k$ is related to a total energy $E = k^2/2$, while the effective "potential energy" $V_{eff}$ is defined as follows:

$$V_{eff}(r) = \frac{h^2}{2r^2}\left(1 - \frac{2M}{r^2}\right) \qquad (34)$$

where $h$ is related to the $\phi$ – angular momentum: $r^2\dot{\phi} = h$ and we consider it as a free parameter. Differentiating (34) with respect to affine-parameter $\sigma$ yields:

$$\ddot{r} = -\frac{dV_{eff}}{dr} = \frac{h^2}{r^3}\left(1 - \frac{4M}{r^2}\right) \qquad (35)$$

To find the related circular motion, consider the condition $\dot{r} = 0$, in which $E = V_{eff}$, such that:

$$E_0 = \frac{h^2}{2r_0^2}\left(1 - \frac{2M}{r_0^2}\right) \qquad (36)$$

with $r_0$ is a constant. From $\ddot{r} = 0$ condition, which is equivalent to $dV_{eff}/dr = 0$, one find the following solution for a specific circular motion with constant $r_0$ which defines the photon sphere:

$$r_0 \equiv r_{PS} = 2\sqrt{M} \qquad (37)$$

that confined in a hyperplane with constant $\theta$ and $\alpha \equiv \alpha_{PS} = 0$. In this condition $E_0 = h^2/16M$.

*4.2. Dynamical System of Photon Motions with Non-Zero $\alpha$ – Angular Momentum*

Now, let us consider the case with $\beta \neq 0$, where we have to take into account the non-zero $\alpha$ – angular momentum condition. To derive the corresponding Newtonian-like orbit equation, we first substitute Eq. (5) into Eq. (6) such that we can recast Eq. (6) into:

$$r^2\dot{\alpha} = \frac{h}{\cos\alpha} \qquad (38)$$

which defines an $\alpha$ – angular momentum. Based on Eq. (38) the effective "potential energy" is now defined as follows:

$$V_{eff}(r;\alpha) = \frac{h^2}{r^2 \cos^2 \alpha}\left(1 - \frac{2M}{r^2}\right) \quad (39)$$

Note that this effective potential energy now depends on the extra spatial dimension related angle $\alpha$. Differentiating (39) with respect to affine-parameter $\sigma$ yields:

$$\ddot{r} = -\frac{dV_{eff}}{dr} = \frac{2h^2}{r^2 \cos^2 \alpha}\left[\frac{1}{r}\left(1 - \frac{4M}{r^2}\right) - \left(1 - \frac{2M}{r^2}\right)\tan\alpha\frac{d\alpha}{dr}\right] \quad (40)$$

Similar to the previous case, by invoking condition $\dot{r} = 0$ yields:

$$E_0 = \frac{h^2}{r_0^2 \cos^2 \alpha_0}\left(1 - \frac{2M}{r_0^2}\right) \quad (41)$$

while from $\ddot{r} = 0$ condition, one finds the following solution:

$$r_0 = 2\sqrt{M}, \; \alpha_0 = 0 \quad (42)$$

in which the related radius coincides with the previous $\beta = 0$ case. Comparing Eq. (10), (38) and (41), it can be identified that for $\beta \neq 0$, $\gamma = \beta/k$ with $k = \sqrt{2E}$, yields

$$h_0 = \frac{\beta \cos \alpha_0}{\sqrt{2E_0}} \text{ and } E_0 = \frac{\beta\sqrt{1 - 2M/r_0^2}}{\sqrt{2r_0}} \quad (43)$$

In contrast with the previous $\beta = 0$ case, in this case, the $h_0$ and $E_0$ parameters depend on $\beta$. For the case $\beta = 8M$ with condition (42), we found from Eq. (43):

$$h_0 = 4\sqrt[4]{M^3} \text{ and } E_0 = 2\sqrt{M} \quad (44)$$

At this point, similar to Eq. (17), in order to understand the meaning of the second time-coordinate singularity in terms of the dynamics of $\alpha(r)$, where based on Eq. (4), (5) and (38) of affine-parameter dependent $\alpha$, the related differential equation is given as follows:

$$\frac{d\alpha}{dr} = \frac{h}{r}\sqrt{\frac{1}{2\Gamma(r,\alpha)}} \quad (45)$$

with $\Gamma(r,\alpha) = r^2 \cos^2\alpha \left[E - V_{eff}\right] > 0$. Clearly, the condition of $E = V_{eff}$ leads Eq. (45) ill-defined. In the meantime, it is readily seen that for the specific conditions given by Eq. (43), which found from the invoking: $\dot{r} = \ddot{r} = 0$ condition, the condition $\Gamma(r,\alpha) = 0$ is valid for all $\beta \neq 0$ cases, in which the related radiuses given by Eq. (12) lead to the second time-coordinate singularities. In other words, the occurrence of the related time-coordinate of photon motion is associated with a specific condition of angular-momentum and energy parameters. It will be demonstrated in the next discussion that the ill-defined condition of Eq. (45) under Eq. (43) forbids the existence of additional photon sphere.

Comparing Eq. (17) with Eq. (45), it is now understood that the motion of a photon with (i) zero [non-zero] $\phi - [\alpha -]$ angular momentum, and (ii) its motion around photon sphere with non-zero [non-zero] $\phi - [\alpha -]$ angular momentum are related to a different $\alpha(r)$ function, where for a photon in a such motion (i) experiences the corresponding photon sphere as a time-coordinate singularity.

To elaborate further the related photon motion, we consider to transform Eq. (40) into a first order ordinary differential system in terms of dynamical system approach. We consider the dynamics of $r$ as a function time-coordinate $t$, in which based on Eq. (3) and (40) it can be written in the following forms:

$$\frac{dr_1}{dt} = \frac{r_2}{\sqrt{2E}}\left(1 - \frac{2M}{r_1^2}\right) \tag{46}$$

$$\frac{dr_2}{dt} = \frac{2h^2}{\sqrt{2E}r_1^3 \cos^2\alpha}\left(1 - \frac{2M}{r_1^2}\right)\left[\left(1 - \frac{4M}{r_1^2}\right) - \frac{h\tan\alpha}{\sqrt{2\Gamma(r_1,\alpha)}}\left(1 - \frac{2M}{r_1^2}\right)\right] \tag{47}$$

Here, we have implicitly transformed $r_1 \equiv r(\sigma)\left[r_2 \equiv \dot{r}(\sigma)\right] \rightarrow r_1 \equiv r(t)\left[r_2 \equiv r_2(t)\right]$, wherein this coordinate $r_2 \neq dr_1/dt$, but it does relate to the slope of $r_1$. The explicit expression for the initial condition of $r_2$ will be shown later [see Eq. (52)]. The function $\Gamma(r_1, \alpha)$ is given by Eq. (45). In the meantime, the dynamics of $\phi(t)$ and $\alpha(t)$ functions are governed by the following differential equations:

$$\frac{d\phi}{dt} = \frac{h}{\sqrt{2E}r_1^2 \cos^2\alpha}\left(1 - \frac{2M}{r_1^2}\right) \quad (48)$$

$$\frac{d\alpha}{dt} = \frac{h}{\sqrt{2E}r_1^2 \cos\alpha}\left(1 - \frac{2M}{r_1^2}\right) \quad (49)$$

respectively, which are constructed from Eq. (3), (5) and (38), where $h$ and $E$ are free parameters. Note that for the case of $\beta = 0$, based on Eq. (35), the related differential equation for $r_2(t)$ and $\phi(t)$ functions are reduced into the following forms:

$$\frac{dr_2}{dt} = \frac{h^2}{\sqrt{2E}r_1^3}\left(1 - \frac{2M}{r_1^2}\right)\left(1 - \frac{4M}{r_1^2}\right) \quad (50)$$

$$\frac{d\phi}{dt} = \frac{h}{\sqrt{2E}r_1^2}\left(1 - \frac{2M}{r_1^2}\right) \quad (51)$$

respectively, which are free from $\alpha$.

At this point, now it becomes clear from Eq. (47) that unlike the case with $\beta = 0$, the case with $\beta \neq 0$ does not constitute photon sphere due to the ill-defined condition of $\Gamma(r_1, \alpha)$ [see Eq. (45)] and the time-coordinate dependent of $\alpha$, where its dynamics is given by Eq. (49).

Recalling Eq. (33), which is also valid for the case $\beta \neq 0$, the expression of $\dot{r}$ can be recast into the following time-coordinate dependent function:

$$\frac{dr}{dt} = \sqrt{\left(1 - \frac{V_{eff}}{E}\right)\left(1 - \frac{2M}{r^2}\right)} \quad (52)$$

where from Eq. (34) and (39), the effective potential function $V_{eff}$ for the case of $\beta = 0$ and $\beta \neq 0$ are given by $V_{eff} = h^2\left(1 - 2M/r_1^2\right)/2r_1^2$ and $V_{eff} = h^2\left(1 - 2M/r_1^2\right)/r_1^2\cos^2\alpha$, respectively. Clearly from Eq. (52) that the condition $E > V_{eff}$ should be satisfied, consistent with the reality condition of $\Gamma(r,\alpha)$ given by Eq. (45). It is interesting to note that by equating Eq. (52) and Eq. (46) yields the expression:

$$\dot{r}_2 = \sqrt{2(E - V_{eff})} \tag{53}$$

which is nothing but a function of the affine-parameter $[\dot{r}_2 \equiv \dot{r}_2(\sigma)]$ related to Eq. (31). For later numerical calculation, the expression (53) can be used to determine the initial condition of $r_2(t)$ in Eq. (46), because it is reasonable to assume that $\dot{r}_2(\sigma = 0) = r_2(t = 0)$.

To study the dynamics of photon motion around the photon sphere based on dynamical system approach in terms of time-coordinate, we consider first the case $\beta = 0$. It is easy to prove that there are two initial condition critical points admitted by Eq. (46) and (50). The first critical point is given by: $(r_{0,1}^{(1)}, r_{0,2}^{(1)}) = (\sqrt{2M}, r_{0,2})$, where for the sake of generality, the parameter $r_{0,2}$ is considered arbitrary, i.e., $r_{0,2} \equiv r_2(t = 0)$. In the meantime, the second critical point is given by: $(r_{0,1}^{(2)}, r_{0,2}^{(2)}) = (2\sqrt{M}, 0)$. Note that all these critical points are related to the radius of the event horizon and the photon sphere, respectively.

Linearizing Eq. (46) and (50) around their first critical point yield the following differential equations:

$$\begin{pmatrix} dr_1/dt \\ dr_2/dt \end{pmatrix} = \begin{pmatrix} r_{0,2}/\sqrt{EM} & 0 \\ -h^2/2\sqrt{2E}M^2 & 0 \end{pmatrix} \begin{pmatrix} r_1 \\ r_2 \end{pmatrix} \tag{54}$$

with eigenvalues:

$$\vec{\lambda} = \left(r_{0,2}/\sqrt{EM}, 0\right) \tag{55}$$

Here, $E$ and $h$ are free parameters, including the energy condition given by Eq. (36). It is readily seen that the eigenvalues (55) are related to an unstable node with negative attractor for, in general,

$r_{0,2} >$. Nevertheless, for the case $r_{0,2} \to 0$, it will be shown later numerically that it is related to a stable node with positive attractor. Meanwhile, for the second critical point we obtain:

$$\begin{pmatrix} dr_1/dt \\ dr_2/dt \end{pmatrix} = \begin{pmatrix} 0 & 1/2\sqrt{2E} \\ h^2/16\sqrt{2E}M^2 & 0 \end{pmatrix} \begin{pmatrix} r_1 \\ r_2 \end{pmatrix} \quad (56)$$

where the corresponding eigenvalues are given as follows:

$$\vec{\lambda} = \left( h/8\sqrt{E}M, -h/8\sqrt{E}M \right) \quad (57)$$

which clearly indicates that this second initial condition critical point is an unstable saddle point for $r_{0,2} = 0$. In the meantime, one can guess that the first critical point it admits a stable node point with positive attractor. We discuss further the behavior of the first critical point later based on numerical calculation.

Based on the same analysis implemented to Eq. (46), (47) and (49) for the case $\beta \neq 0$, we observed slightly different dynamical system characteristics. Similar to the previous case, the first critical point admitted for this case is given by: $\left( r_{0,1}^{(1)}, r_{0,2}^{(1)}, r_{0,3}^{(1)} \right) = \left( \sqrt{2M}, r_{0,2}, \alpha_0 \right)$. The parameters $r_{0,2}$ and $\alpha_0$ are also considered arbitrary, i.e., $r_{0,2} \equiv r_2(t=0)$ and $\alpha_0 \equiv \alpha(t=0)$. The linearization form of Eq. (46), (47) and (49) around this critical point yields:

$$\begin{pmatrix} dr_1/dt \\ dr_2/dt \\ d\alpha/dt \end{pmatrix} = \begin{pmatrix} r_{0,2}/\sqrt{EM} & 0 & 0 \\ -h^2/\sqrt{2E}M^2 \cos^2 \alpha_0 & 0 & 0 \\ h/2\sqrt{EM^3} \cos \alpha_0 & 0 & 0 \end{pmatrix} \begin{pmatrix} r_1 \\ r_2 \\ \alpha \end{pmatrix} \quad (58)$$

with eigenvalues:

$$\vec{\lambda} = \left( r_{0,2}/\sqrt{EM}, 0, 0 \right) \quad (59)$$

Here, $E$ and $h$ are free parameters, excluding the angular-momentum and energy conditions given by Eq. (43), since they lead to the ill-defined condition of Eq. (47) with $\Gamma(r, \alpha) = 0$. As mentioned previously, this condition is consistent with the second time-coordinate singularity experienced by a photon with non-zero angular momentums.

Unlike the previous first critical point which has three components, for the second critical point in this $\beta \neq 0$ case we do our dynamical system analysis by not considering the dynamics of $\alpha$, since Eq. (49) has no related critical point. Therefore, the second critical point for this case have only two components which is given by: $\left(r_{0,1}^{(2)}, r_{0,2}^{(2)}\right) = \left(2\sqrt{M}, 0\right)$, such that the linearization form of coupled Eq. (46) and (47) is given as follows:

$$\begin{pmatrix} dr_1/dt \\ dr_2/dt \end{pmatrix} = \begin{pmatrix} 0 & 1/2\sqrt{2E} \\ h^2/8\sqrt{2E}M^2 \cos^2 \alpha_0 & 0 \end{pmatrix} \begin{pmatrix} r_1 \\ r_2 \end{pmatrix} \tag{60}$$

with eigenvalues:

$$\vec{\lambda} = \left( h/4\sqrt{2E}M \cos\alpha_0, -h/4\sqrt{2E}M \cos\alpha_0 \right) \tag{61}$$

Here, we assume that $\alpha_0 \equiv \alpha(t=0)$. From eigenvalues (59) and (61) one finds that the first critical point is likely exhibits an unstable node point with negative attractor, whereas the second critical point is an unstable saddle point for $r_{0,2} = 0$ case.

Based on Eq. (43), we are now at a position to analyze the case $\beta \neq 0$. First, let us consider the inversion of Eq. (15) and treat it as a function of $r = r_{0,1}$ as given as follows:

$$\beta \equiv \beta(r_{0,1}) = \frac{r_{0,1}^4}{r_{0,1}^2 - 2M} \tag{62}$$

It is easy to prove that $\beta(r_{0,1} = r_{PS}) = 8M$ and $\beta(r_{0,1} \to r_{EH}) \to \infty$, such that $8M \leq \beta < \infty$. From Eq. (43) and (62), it is clear that the motion with $\beta \geq 8M$ is related to a photon with higher energy and momentum. Based on this condition, one can fix a certain radius $r = r_{0,1}$, determines $\beta$ from Eq. (62), and then solve the Eq. (46)-(49) numerically with $\alpha_0 = \alpha_{PS} = 0$. Meanwhile, it should be re-emphasized that the condition (43) leads Eq. (45) become singular for all initial conditions where $\Gamma(r_{0,1},\alpha_0)=0$. In other words, as previously mentioned, the time-coordinate singularities exist at the higher energy level of energy $\beta \geq 8M$, but does not constitute another photon sphere except at $r_{0,1}=r_{PS}=2\sqrt{M}$ with $\beta=0$. This phenomenon is consistent with the results reported in refs. [24, 25] which stated that no bound circular motions are admitted for a spacetime geometry with spatial dimension more than three. It is still worth to note that the slope of photon trajectory for zero [non-zero] $\phi-[\alpha-]$ angular momentum case is given by Eq. (12). Therefore, in principle, once the related integral can be solved, one can construct the associated Eddington-Finkelstein-like coordinate similar to the previously discussed case with $\beta=8M$.

*4.3. Characteristics of Photon Bending Motions*

One of the important types of photon motions is the bending motion of escaping photons [37], which helps detect the presence of a black hole [1]. We will show in this section that our following numerical simulation result on the bending motion of escaping photon can probably be considered to probe the existence of the extra spatial dimension.

To study the corresponding photon bending motion characteristics, let us first simulate the case $\beta=0$ namely by solving Eq. (46) and (50) numerically using the Runge-Kutta method. We assume that the free parameter $h$ is the same as for $\beta=8M$ case, i.e., $h=h_0=4\sqrt[4]{M^3}$ [see Eq. (44)]. In order to visualize the corresponding photon circular motion at the photon sphere, we represent its dynamics in $(r\cos\phi, r\sin\phi)$ projection plane, in which $r \equiv r_1$ in our dynamical system analysis. Note that this chosen projection plane can be viewed as a projection to the equator of a $(3+1)$–dimensional spacetime. It is readily seen from Eq. (53), that we can set the initial condition $r_{0,2}$ for $\beta=0$, in which the condition $E \geq V_{eff}$ should be satisfied.

Depicted in Fig. 2a is the motion of photon with $\beta=0$ with $r_{0,2}=0$, where the associated $\phi-$ angular momentum is a constant, i.e., $r_{ph}^2 d\phi/dt = h/2\sqrt{2E}$ exhibiting a periodic behavior with

$d\phi/dt = h/8\sqrt{2ME}$. Here, the initial position of the photon is initially located at the photon sphere with $r_{0,1} = \sqrt{4M} = r_{PS}$. Varying the related initial position, i.e., $r_{0,1} = \sqrt{4.4M} > r_{PS}$ leads the photon to escape from the photon sphere as shown in Fig. 2b for $r_{0,2} \to 0$, i.e., $r_{0,2} = 0.1$. In contrast to $r_{0,1} = r_{PS}$ case, the related $\phi$-angular momentum: $\lim_{t \to \infty} r^2 d\phi/dt \to h/\sqrt{2E}$, with $\lim_{t \to \infty} d\phi/dt \to 0$. This non-zero angular momentum means that the photon trajectory exhibits bending motion. On the other hand, for $r_{0,1} = \sqrt{3.6M} < r_{PS}$ the photon is being trapped into the event horizon such that it cannot be detected. Our numerical result indicate that the related critical point is a stable node with positive attractor for $r_{0,2} \to 0$, depending on how close the initial condition $r_{0,1}$ to $r_{PS}$.

We restrict our next simulations only to the cases of an outgoing photon which is departed from the initial position at or near the photon sphere. Exemplified in Fig. 3a are the photon trajectories for the case with $\beta = 8M$, which are simulated by solving Eq. (46)-(49) numerically, with initial conditions $\alpha_0 = \alpha_{PS} = 0$, $r_{0,1} = r_{PS}$ and $r_{0,2} = 1.7321$. In this case, the corresponding $\phi$-angular momentums satisfy $\lim_{t \to \infty} r^2 d\phi/dt \to h/\sqrt{2E} \cos\alpha$, with $\lim_{t \to \infty} d\phi/dt \to 0$. Similar to the previous case with $\beta = 0$, it is shown that the photon is being "pushed" away from the photon sphere. Demonstrated in Fig. 3a that the photon trajectory for this case is more bent than the previous $\beta = 0$ case, and can be explained as a consequence of a non-zero $\alpha$-angular momentum. Unlike the $\beta = 0$ case, we found that there is no photon sphere of circular motion for $r_{0,2} = 0$ with $E = V_{eff}$ in this case as expected. As mentioned previously, this condition can be understood from the fact that $\Gamma(r_1, \alpha)$ in the Eq. (45) becomes zero.

As the last example, we consider the case $\beta > 8M$, i.e., $\beta = 9M$ with the initial conditions are set to: $r_{0,1} = \sqrt{6M}$ and $r_{0,2} = 1.7321$. Depicted in Fig. 3b is the related trajectory with the same $\phi$-angular momentum as the previous $\beta = 8M$ case. Compared to the $\beta = 8M$ case in Fig. 3a, as expected, it is shown that the corresponding trajectory is less bent since it has higher energy level.

To this end, it is interesting to point out that the dynamics of the photon bending motion is strongly affected by the existence of the extra spatial dimension represented by $\alpha$ – coordinate lying in a bulk universe and hidden to us. From our numerical simulation represented in the related projection plane, it is revealed that the motions with $\beta \neq 0$ exhibit more bent trajectories compared to the $\beta = 0$ case. Based on this finding, it is suggested that the characteristics of a photon bending motion around a black hole with strong gravitational field can be used to probe the related extra spatial dimension existence [35, 38], in addition to the previously proposed probing mechanisms based on the massive photon model [39] and gravitational waves [40]. Indeed, this remarkable feature needs further study in order to investigate its astrophysical consequences and it should be considered as an early indication for such possible detection mechanism. The shadows of M87* [35] and Sagittarius A* [41] black holes should be a good playground to test whether the dynamics of photons around them are obeying the potential given by Eq. (34) instead of the $(3+1)$– dimensional potential, e.g., [36] since we cannot effectively distinguish the photon dynamics in the strong gravitational field around a black hole in either four- or five-dimensional gravitational theories for the current observation, e.g., [35]. However, it should be emphasized that the weak-gravitational-field related Newtonian potential in our formulation, which is proportional to $1/r^2$, does not appropriate with the observational result [42]. Furthermore, it is also interesting to study cases with spatial dimensions of more than four. Although it is intriguing to be investigated, but we expect to find similar characteristics.

## 5. Conclusion

We have discussed the photon motion in a $(4+1)$– dimensional Schwarzschild black hole admitting a rotational isometry metric. We have constructed the Eddington-Finkelstein-like coordinate related to a photon motion with zero $\phi-$angular momentum. We found that for a non-zero $\alpha-$angular momentum case, the corresponding time-coordinate admits a secondary time-coordinate singularity in addition to the singularity at the event horizon. By examining the circular motion of a photon with a constant radius based on the dynamical system approach, we have identified that the secondary time-coordinate singularity is nothing but related to a photon sphere experienced by a photon that moves with non-zero [zero] $\phi-[\alpha-]$angular momentum. We have shown that the higher-order time-coordinate singularities are related to the photons with specific angular-momentum and energy levels, but they do not lead to any additional photon spheres. Based on the numerical calculation, we have also discussed the simulation results regarding a photon dynamic around the associated photon sphere. The results suggest that the trajectory characteristic of photon bending motion can probably be considered to be used as a probe to detect the existence of the extra spatial dimension represented by the $\alpha-$coordinate.

**Conflict of Interest**

The authors declare that there is no conflict of interest.

**Acknowledgment**

This work is funded by the WCR grant 2021 from the National Research and Innovation Agency (BRIN) of Indonesia, under contract no. 2348/IT3.L1/PN/2021. This work is also partially funded by Riset ITB 2019 grant from Institut Teknologi Bandung. While the work of BEG is also partially supported by P3MI ITB 2020 grant from Institut Teknologi Bandung.

**References**

1. The Event Horizon Telescope Collaboration et al, First M87 Event Horizon Telescope Results. I. The Shadow of the Supermassive Black Hole, The Astrophysical Journal Letters **875** (2019) L1.

2. K. Schwarzschild (translation and foreword by S. Antoci and A. Loinger), On the Gravitational Field of a Mass Point According to Einstein's Theory, arXiv:physics/9905030v1.

3. J. P. Luminet, Black Holes: A General Introduction. In: Hehl F., Kiefer C., Metzler R. (eds) Black Holes: Theory and Observation. Lecture Notes in Physics, **514**. Springer, Berlin, Heidelberg (1998).

4. A. Zee, Einstein Gravity in a Nutshell, Princeton University Press, USA, 2013, pp. 362 – 375.

5. R. P. Kerr, Gravitational Field of a Spinning Mass as an Example of Algebraically Special Metrics, Physical Review Letters **11** (1963) 237.

6. S. Abdolrahimi, V. P. Frolov, A. A. Shoom, Interior of a Charged Distorted Black Hole, Physical Review D **80** (2009) 024011.

7. H. M. Siahaan, Triyanta, Semiclassical Methods for Hawking Radiation from a Vaidya Black Hole, International Journal of Modern Physics A **25** (2010) 145.

8. H. M. Siahaan, Destroying Kerr-Sen Black Holes, Physical Review D **93** (2016) 064028.

9. W. Chen, H. Lü, C. N. Pope, General Kerr–NUT–AdS Metrics in All Dimensions, Classical and Quantum Gravity **23** (2006) 5323.

10. T. Y. Yu, W. Y. Wen, Cosmic Censorship and Weak Gravity Conjecture in the Einstein–Maxwell-Dilaton Theory, Physics Letters B **781** (2019) 713.

11. B. Gwak, Thermodynamics and Cosmic Censorship Conjecture in Kerr-Newman-de Sitter Black Hole, Entropy **20** (2019) Art. No. 855.

12. K. Belczynski, G. Wiktorowicz, C. L. Fryer, D. E. Holz, V. Kalogera, Missing Black Holes Unveil the Supernova Explosion Mechanism, The Astrophysical Journal **757** (2012) 91.

13. A. Övgün, I. Sakalli, J. Saavedra, Quasinormal Modes of a Schwarzschild Black Hole Immersed in an Electromagnetic Universe, Chinese Physics C **42** (2018) 105102

14. P. T. Chruściel, J. L. Costa, M. Heusler, Stationary Black Holes: Uniqueness and Beyond, Living Review in Relativity **15** (2012) 7.

15. M. Z. Iofa, Near-Horizon Symmetries of the Schwarzschild Black Holes with Supertranslation Field, Physical Review D **99** (2019) 064052.

16. M. Eune, W. Kim, Test of Quantum Atmosphere in the Dimensionally Reduced Schwarzschild Black Hole, Physics Letters B **798** (2019) 135020.

17. E. Greenwood, Classical and Quantum Equations of Motion of a 4-Dimensional Schwarzschild–Ads and Reissner–Nordström–Ads Black Hole, International Journal of Modern Physics D **28** (2019) 1950061.

18. X. B. Huang, Unification of Gravitation, Gauge Field and Dark Energy, International Journal of Modern Physics A **21** (2006) 1341.

19. C. Deffayet, G. Dvali, G. Gabadadze, Accelerated Universe from Gravity Leaking to Extra Dimensions, Physical Review D **65** (2002) 044023.

20. A. Lue, G. Starkman, Gravitational Leakage into Extra Dimensions: Probing Dark Energy Using Local Gravity, Physical Review D **67** (2003) 064002.

21. K. Ghosh, D. Karabacak, S. Nandi, Universal Extra Dimension Models with Gravity Mediated Decays After LHC Run II Data, Physics Letters B **788** (2019) 388.

22. J. M. Overduin, P. S. Wesson, Kaluza-Klein Gravity, Physics Report **283** (1997) 303.

23. F. R. Tangherlini, Schwarzschild Field in $n-$Dimensions and the Dimensionality of Space Problem, Il Nuovo Cimento **27** (1963) 636.

24. R. Emperan, H. S. Reall, Black Holes in Higher Dimensions, Living Review in Relativity **11** (2008) 6.

25. M. Arik, D. Çiftci, Schwarzschild-de Sitter Black Holes in $(4+1)-$Dimensional Bulk, Modern Physics Letters A **22** (2007) 289.

26. N. Dadhich, Bound Orbits and Gravitational Theory, Physical Review D **88** (2013) 124040.

27. M. Bhattacharya, N. Dadhich, B. Mukhopadhyay, Study of Motion Around a Static Black Hole in Einstein and Lovelock Gravity, Physical Review D **91** (2015) 064063.

28. R. Emperan, H. S. Reall, Black Rings, Classical and Quantum Gravity **23** (2006) R169.

29. F. D. Mazzitelli, The Infalling Photon, the Infalling Particle, and the Observer at Rest Near the Horizon of a Black Hole, European Journal of Physics **41** (2020) 065601

30. F. Atamurotov, B. Ahmedov, A. Abdujabbarov, Optical Properties of Black Holes in the Presence of a Plasma: The Shadow, Physical Review D **92** (2015) 084005

31. E. Teo, Spherical Photon Orbits Around a Kerr Black Hole, General Relativity and Gravitation **35** (2003) 1909

32. M. Jamila, S. Hussain, B. Majeed, Dynamics of Particles Around a Schwarzschild-like Black Hole in the Presence of Quintessence and Magnetic Field, European Journal of Physics C **75** (2015) 24.

33. A. Nazara, S. Hussain, A. Aslam, T. Hussain, M. Ozair, Charged Particle Dynamics in the Vicinity of Black Hole from Vector-tensor Theory of Gravity Immersed in an External Magnetic Field, Results in Physics **14** (2019) 102418.

34. M. Sharif, S. Iftikhar, Dynamics of Particles Near Black Hole with Higher Dimensions, European Physical Journal C **76** (2016) 404.

35. S. Vagnozzi and L.Visinelli, Hunting for Extra Dimensions in the Shadow of M87*, Physical Review D **100** (2019) 024020.

36. M.P. Hobson, G. P. Efstathiou, A. N. Lasenby, General Relativity: An Introduction for Physicists, Cambridge University Press, Cambridge, UK, 2006, pp 379 – 380.

37. S. V. Iyer, A. O. Petters, Light's Bending Angle due to Black Holes: from the Photon Sphere to Infinity, General Relativity & Gravitation **39** (2007) 1563.

38. A. Chatterjee, S. K. Chakrabarti, H. Ghosh, Images and Spectral Properties of Two-Component Advective Flows Around Black Holes: Effects of Photon Bending, Monthly Notices of the Royal Astronomical Society **465** (2017) 3902.

39. G. Alencar, C.R. Muniz, R.R. Landim, I.C. Jardim, R.N. Costa Filho, Photon Mass as a Probe to Extra Dimensions, Physics Letters B **759** (2016) 138.

40. Y. Du, S. Tahura, D. Vaman, K. Yagi, Probing Compactified Extra Dimensions with Gravitational Waves, Physical Review D **103** (2020) 044031.

41. Z. Zhu, M. D. Johnson, R. Narayan, Testing General Relativity with the Black Hole Shadow Size and Asymmetry of Sagittarius A*: Limitations from Interstellar Scattering, The Astrophysical Journal **870** (2019) 6.

42. C. Rothleitner, Ultra-Weak Gravitational Field Detected, Nature **591** (2021) 209.

**FIGURE CAPTIONS**

Fig. 1. Photon motion of non-zero $\alpha$ – angular momentum exhibiting the two coordinate singularities at event horizon with $r = \sqrt{2M}$ and photon sphere with $r = 2\sqrt{M}$ for ingoing (solid curves) and outgoing (dash curve) motions.

Fig. 2. Photon motion with $\beta = 0$ for (a) $r_{0,2} = 0$ with $r_{0,1} = \sqrt{4M} = r_{PS}$ (solid-black curve) (b) $r_{0,2} = 0.1$ with $r_{0,1} = \sqrt{3.6M} < r_{PS}$ (solid-black curve) and $r_{0,1} = \sqrt{4.4M} > r_{PS}$ (dotted-black curve). The red and blue dash circles are the event horizon and photon sphere, respectively.

Fig. 3. Photon motion with $r_{0,2} = 1.7321$ for (a) $\beta = 8M$ at $r_{0,1} = r_{ph} = \sqrt{4M}$ (b) $\beta = 9M$ at $r_{0,1} = \sqrt{6M}$ (solid-black curves). All cases are compared with the motion case $\beta = 0$ (dotted-black curve). The red and blue dash circles are the event horizon and photon sphere, respectively.

Figure 1

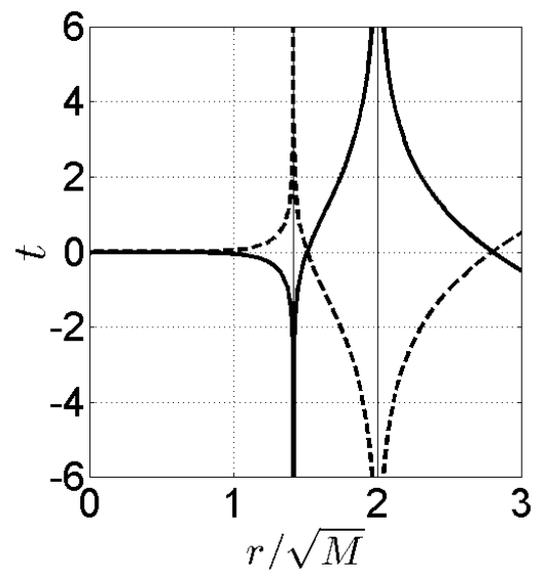

Figure 2

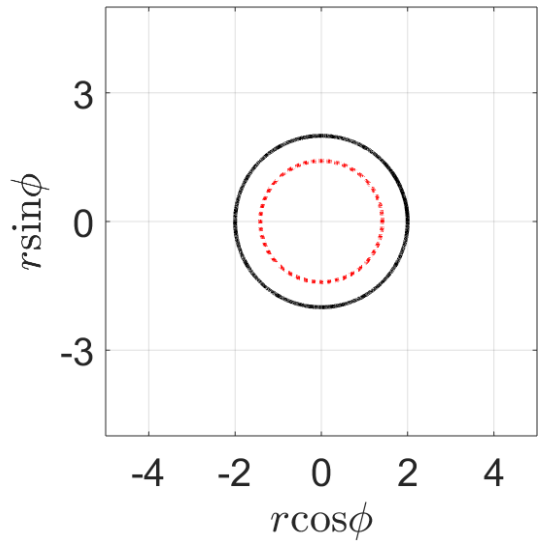
(a)

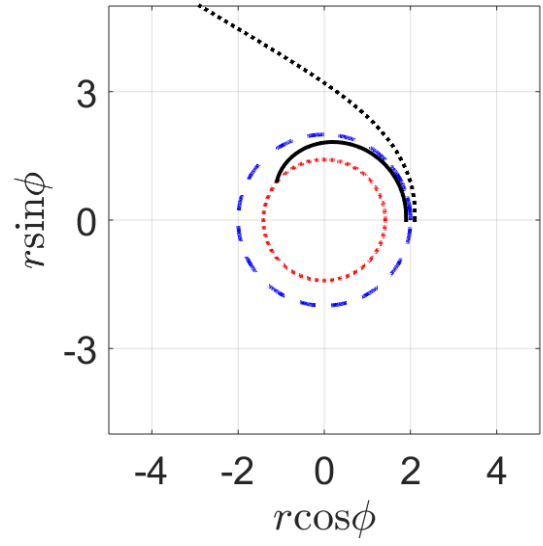
(b)

Figure 3

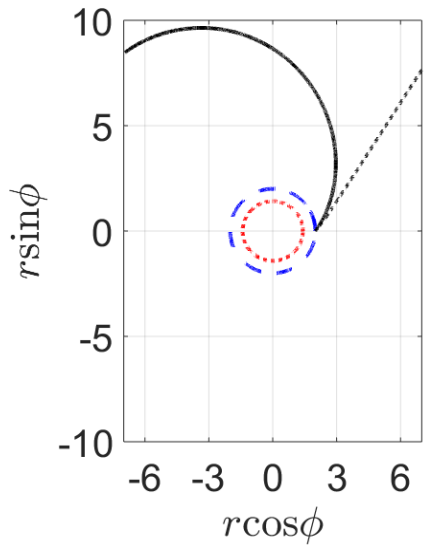
(a)

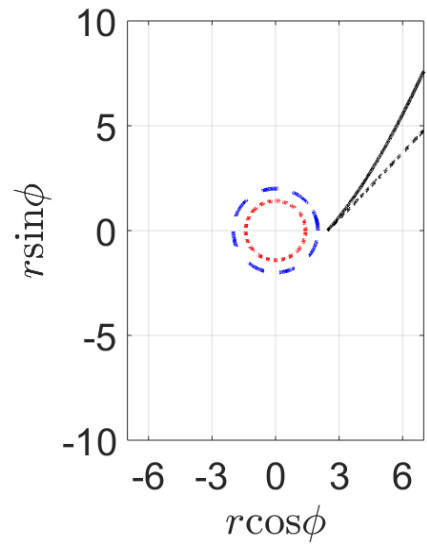
(b)